\documentclass[prl,aps,twocolumn,superscriptaddress,amsmath,amssymb]{revtex4}
\usepackage{graphicx}
\usepackage{dcolumn}
\usepackage{bm}
\usepackage{color}

\begin{document}

\title{Common Origin of the Circular-dichroism Pattern in ARPES of 
SrTiO$_3$ and Cu$_x$Bi$_2$Se$_3$}

\author{Y.~Ishida}
\affiliation{ISSP, University of Tokyo, Kashiwa, 
Chiba 277-8561, Japan}

\author{H.~Kanto}
\affiliation{ISSP, University of Tokyo, Kashiwa, 
Chiba 277-8561, Japan}

\author{A.~Kikkawa}
\affiliation{Correlated Electron Research Group (CERG), 
RIKEN Advanced Science Institute, Wako 351-0198, Japan}

\author{Y.~Taguchi}
\affiliation{Correlated Electron Research Group (CERG), 
RIKEN Advanced Science Institute, Wako 351-0198, Japan}
\affiliation{Cross-Correlated Materials Research Group (CMRG), 
RIKEN Advanced Science Institute, Wako 351-0198, Japan}

\author{Y.~Ito}
\affiliation{ISSP, University of Tokyo, Kashiwa, 
Chiba 277-8561, Japan}

\author{Y.~Ota}
\affiliation{ISSP, University of Tokyo, Kashiwa, 
Chiba 277-8561, Japan}

\author{K.~Okazaki}
\affiliation{ISSP, University of Tokyo, Kashiwa, 
Chiba 277-8561, Japan}

\author{W.~Malaeb}
\affiliation{ISSP, University of Tokyo, Kashiwa, 
Chiba 277-8561, Japan}

\author{M.~Mulazzi}
\affiliation{ISSP, University of Tokyo, Kashiwa, 
Chiba 277-8561, Japan}

\author{M.~Okawa}
\affiliation{ISSP, University of Tokyo, Kashiwa, 
Chiba 277-8561, Japan}

\author{S.~Watanabe}
\affiliation{ISSP, University of Tokyo, Kashiwa, 
Chiba 277-8561, Japan}

\author{C.-T.~Chen}
\affiliation{Technical Institute of Physics and Chemistry, 
Chinese Academy of Science, Zhongguancun, Beijing 100080, China}

\author{M.~Kim}
\affiliation{Department of Advanced Materials Science, University of 
Tokyo, Kashiwa, Chiba 277-8561, Japan}

\author{C.~Bell}
\affiliation{Department of Advanced Materials Science, University of 
Tokyo, Kashiwa, Chiba 277-8561, Japan}
\affiliation{Department of Applied Physics and Stanford Institute for 
Materials and Energy Science, Stanford University, Stanford, California 
94305, USA}

\author{Y.~Kozuka}
\affiliation{Department of Advanced Materials Science, University of 
Tokyo, Kashiwa, Chiba 277-8561, Japan}

\author{H.~Y.~Hwang}
\affiliation{Correlated Electron Research Group (CERG), 
RIKEN Advanced Science Institute, Wako 351-0198, Japan}
\affiliation{Department of Advanced Materials Science, University of 
Tokyo, Kashiwa, Chiba 277-8561, Japan}
\affiliation{Department of Applied Physics and Stanford Institute for 
Materials and Energy Science, Stanford University, Stanford, California 
94305, USA}

\author{Y.~Tokura}
\affiliation{Correlated Electron Research Group (CERG), 
RIKEN Advanced Science Institute, Wako 351-0198, Japan}
\affiliation{Cross-Correlated Materials Research Group (CMRG), 
RIKEN Advanced Science Institute, Wako 351-0198, Japan}
\affiliation{Department of Applied Physics, University of Tokyo, 
Tokyo 113-8656, Japan}

\author{S.~Shin}
\affiliation{ISSP, University of Tokyo, Kashiwa, 
Chiba 277-8561, Japan}
\affiliation{CREST, Japan Science and Technology Agency, 
Tokyo 102-0075, Japan}

\date{\today}

\begin{abstract}
Circular dichroism in the angular distribution (CDAD) 
of photoelectrons 
from SrTiO$_3$:Nb and Cu$_x$Bi$_2$Se$_3$ 
is investigated by 7-eV laser ARPES. 
In addition to the well-known node that occurs in CDAD 
when the incidence plane 
matches the mirror plane of the crystal, we show that another type of 
node occurs when the mirror plane of the crystal is vertical to the 
incidence plane and the electronic state is two dimensional. 
The flower-shaped CDAD's occurring around the Fermi level of SrTiO$_3$:Nb 
and around the Dirac point of Cu$_x$Bi$_2$Se$_3$ are 
explained on equal footings. 
We point out that the penetration depth of the topological states 
of Cu$_x$Bi$_2$Se$_3$ depends on momentum. 
\end{abstract}

\maketitle 

The interaction of light with matter depends on the polarization 
of the photons. Circular dichroism (CD) is a phenomenon 
in which the response of a system to left and right circularly 
polarized light is different. CD can be microscopically 
attributed to the difference in the material's response 
against opposite helicities of the photons. 
Thus, CD has been actively used for studying magnetic materials 
or those having strong spin-orbit interactions 
\cite{Thole, Carra, Kuch_01}. 
Alternatively, 
left and right circular polarizations are exchanged 
by a mirror operation, and therefore CD is active 
when the measurement breaks symmetry with respect to the reflection; 
i.e., CD occurs when the experimental geometry has ``handedness" 
\cite{Dubs, Schonhense_90}. 

In angle resolved photoemission spectroscopy (ARPES), 
light is shined on a crystal and the energy-and-angle 
distribution  of the photoelectrons is recorded. 
The band structures of crystals and crystal surfaces are 
traced by ARPES, 
and the information is further enriched by 
investigating the CD in the angular distribution (CDAD) of 
the photoelectrons 
\cite{Dubs, Schonhense_90, Kuch_01, Suga, 
Kaminski, Borisenko, Mattia_Ni111, Panaccione, Zabolotnyy, YbRh2Si2}. 
For example, a node in CDAD occurs 
when the incidence plane and the mirror plane of the 
crystal are matched 
\cite{Schonhense_90}. This vertical node, 
which occurs due to reasons of symmetry, has been 
utilized 
in various ARPES studies 
\cite{Kaminski, Borisenko, Mattia_Ni111, Panaccione, Zabolotnyy, YbRh2Si2}.
In this Letter, we show that there is another type of node, 
a {\it horizontal} node, 
which occurs 
due to a combination of the symmetry and dimensionality of the 
initial electronic state. 
We first investigate 
photoemission matrix elements and 
derive the condition for 
the occurrence of the 
horizontal node. 
Then we introduce and derive information from 
the horizontal nodes occurring 
in the CDADs of SrTiO$_3$:Nb and 
Cu$_x$Bi$_2$Se$_3$.

\begin{figure}[htb]
\begin{center}
\includegraphics[width=6.9cm]{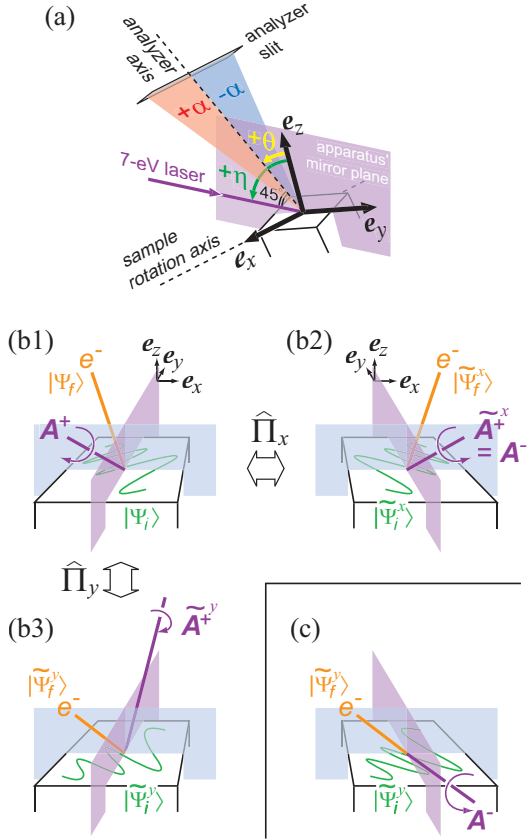}
\caption{\label{fig1} A bird's-eye view of photoemission. 
(a) ARPES geometry. The polarization of the laser is controlled by a 
wave plate \cite{Kiss}. 
(b1) Photoemission by a right circularly polarized photon. (b2) and (b3) 
are mirror reflections of (b1) with respect to the $x=0$ (incidence) 
plane and $y=0$ plane, respectively. 
$\hat{\Pi}_l$ ($l = x$ or $y$) is the operator for 
the reflection with respect to the $l=0$ plane. 
A tilde accompanied by $l$ is attached to 
the reflected states and the reflected vector potentials. 
(c) A configuration 
for achieving photoemission into $|\tilde{\Psi}^y_i\rangle$. 
Note the resemblance of (c) and (b3) up to the direction of the 
rotation of the vector potentials in the $xy$ plane. 
}
\end{center}
\end{figure}

The experimental geometry of 7-eV laser ARPES \cite{Kiss} 
is shown in Fig.\ \ref{fig1}(a). 
By using an orthogonal basis fixed on the 
sample (we take ${\bm e}_x$ along the sample 
rotation axis and ${\bm e}_y$ along 
the sample surface), 
the vector potentials for right ($\bm{A}^+$) and left ($\bm{A}^-$) 
circular polarizations are described as 
$\bm{A}^\pm=\bm{A}^\pm_{\rm pes}e^{-i(\omega t+\varphi)} + {\rm c.c.}$, 
where 
$\bm{A}^\pm_{\rm pes} = 
\frac{A}{2}(-1, \mp i \cos\eta, \mp i\sin\eta)$, 
$\eta$ is the angle 
between the laser beam and ${\bm e}_z$, and $\varphi$ is a phase. 
The analyzer collects photoelectrons emitted within the acceptance 
angle $|\alpha| \textless 18^{\circ}$ ($\alpha=0^{\circ}$ is the direction 
to the analyzer axis), and the photoelectron distribution 
($I$) is 
recorded as functions of $\theta$, $\alpha$, and $E_B$, 
where $\theta$ is the rotation angle of ${\bm e}_z$ with 
respect to the analyzer axis and $E_B$ is the binding energy 
referenced to $E_F$ of gold. The spectra are 
recorded at 10 K with an energy resolution of $\sim$3 meV.

In general, a photoemission event from a given initial-to-final 
state ($|\Psi_i\rangle$ to $|\Psi_f\rangle$) under a given experimental 
setup has the same cross section to another event which is a mirror 
reflection of the 
original one \cite{93_Venus_CoreMCD}. 
Here, everything should be reflected, 
not just the crystal and the incident light, but also 
$|\Psi_{i,f}\rangle$ into 
$|\tilde{\Psi}_{i,f}\rangle \equiv \hat{\Pi}|\Psi_{i,f}\rangle$, 
and even the direction of the circulating currents responsible for magnetism, 
if any. 
We consider a one-step photoemission process, so that 
$|\Psi_f\rangle$ is an inverse low-energy electron diffraction (LEED) 
state extending 
from the sample into the detector 
\cite{Schonhense_90} and 
evaluate the operator under a dipole approximation: 
$I^{\pm} \propto 
|\langle\Psi_f|\bm{A}^{\pm}_{\rm pes}\cdot\hat{\bm{p}}|\Psi_i\rangle|^2$.

When the incidence plane matches the mirror plane 
of the crystal, the result $\tilde{I}^+$ obtained in a reflected 
experiment with respect to the incidence plane 
(this coincides with the apparatus' mirror plane 
in our experimental geometry) 
is the same as that of the original experiment with a 
reversed circular polarization $I^-$, as can be seen 
by comparing Figs.\ \ref{fig1}(b1) 
and \ref{fig1}(b2). 
One can confirm 
$|\langle\Psi_f|\bm{A}^+_{\rm pes}\cdot\hat{\bm{p}}|\Psi_i\rangle| 
= |\langle\tilde{\Psi}^x_f|\bm{A}^-_{\rm pes}\cdot\hat{\bm{p}}|\tilde{\Psi}^x_i\rangle|$ [the left- and right-hand sides of this equation 
correspond to the events 
shown in Fig.\ \ref{fig1}(b1) and \ref{fig1}(b2), respectively] 
with the aid of 
$\hat{\Pi}_x(\bm{A}^+_{\rm pes}\cdot\hat{\bm{p}})\hat{\Pi}_x 
=\bm{A}^+_{\rm pes}\cdot(-\hat{p}_x\bm{e}_x + \hat{p}_y\bm{e}_y + \hat{p}_z\bm{e}_z) 
= \frac{A}{2}(1, -i\cos{\eta}, -i\sin{\eta})\cdot\hat{\bm{p}} 
= -\bm{A}^-_{\rm pes}\cdot\hat{\bm{p}}$. 
Therefore 
$I^+(\theta, \alpha, E_B) = I^-(\theta, -\alpha, E_B)$, 
so that
the angular distribution of the dichroism $I^D = I^+ - I^-$ 
acquires a vertical node, i.e., 
$I^D(\theta, 0^{\circ}, E_B) = 0$.

Next, we consider a case where the mirror plane of the crystal is 
vertical to the incidence plane, and ask whether 
\begin{eqnarray}
I^+(\theta, \alpha, E_B) = I^-(-\theta, \alpha, E_B)
\label{eq1}
\end{eqnarray}
can hold. This corresponds to investigating whether the 
matrix elements for the events shown in 
Figs.\ \ref{fig1}(b1) and \ref{fig1}(c) 
can be equivalent or not.
The incidence angles are different, since the 
laser and the analyzer are fixed in space; thus, 
these two photoemission events 
cannot overlap by any symmetry operations. 
Therefore Eq.\ (\ref{eq1}) does not hold globally. 
Nevertheless, the events shown in 
Figs.\ \ref{fig1}(b3) and \ref{fig1}(c) 
[the former is a reflection of Fig.\ \ref{fig1}(b1) 
with respect to the $zx$ mirror plane, and hence, equivalent to 
the event of Fig.\ \ref{fig1}(b1)] resemble each other: 
In both cases, the initial and final states are the same, and the 
in-plane ($xy$) components of $\bm{A}$ rotate anticlockwise on the 
sample surface. 
Explicitly, Eq.\ (\ref{eq1}) is equivalent to 
\begin{eqnarray}
\lefteqn{
|\langle\tilde{\Psi}^y_f|\left(
\begin{array}{c} -1 \\ i\cos\eta \\ -i\sin\eta \end{array}
\right)
\cdot
\bm{\hat{p}}
|\tilde{\Psi}^y_i\rangle| 
}
\nonumber \\
&=&	
|\langle\tilde{\Psi}^y_f|\left(
\begin{array}{c} -1 \\ i\cos(\eta+2\theta) \\ i\sin(\eta+2\theta) \end{array}
\right)
\cdot
\bm{\hat{p}}
|\tilde{\Psi}^y_i\rangle|,
\label{eq2}
\end{eqnarray}
and the main difference occurs in the sign (phase) 
of the $z$ component of the vector potential with respect to 
the $x$ and $y$ components. 
Thus, when 
\begin{eqnarray}
|\langle\Psi_f|\hat{p}_{z}|\Psi_i\rangle|
\ll|\langle\Psi_f|\hat{p}_{x,y}|\Psi_i\rangle|
\label{eq3}
\end{eqnarray}
is fullfilled, 
Eq.\ (\ref{eq2}) and, hence Eq.\ (\ref{eq1}) holds 
at $\theta\,\sim\,0^{\circ}$, resulting in 
a horizontal node $I^D(0^{\circ}, \alpha, E_B)$ = 0.

\begin{figure}[htb]
\begin{center}
\includegraphics[width=7.5cm]{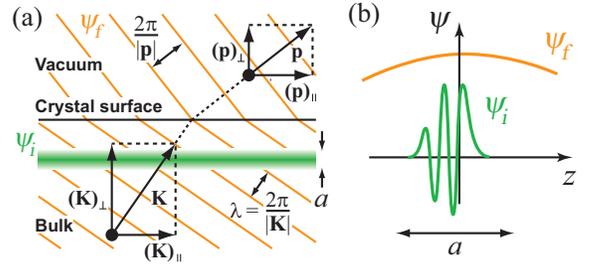}
\caption{\label{fig2} 
Photoemission from a 2D state into an inverse 
LEED state. (a) The 2D and inverse LEED states. Orange 
lines show wave fronts that are refracted at the surface. 
(b) The wave functions of the 2D and inverse LEED states. }
\end{center}
\end{figure}

The condition (\ref{eq3}) is fulfilled  
when $\Psi_i$ is two-dimensional (2D) and spatially confined 
in the $z$ direction 
within a length scale $a$ 
shorter than the de Broglie wave length 
$\lambda$ 
of the photoelectron 
final state, 
as shown in Fig.\ \ref{fig2}. 
Then, 
$\int^{\infty}_{-\infty}\!dz\,\Psi^*_f(x,y,z)\frac{\hbar}{i}
\frac{\partial}{\partial z}\Psi_i(x,y,z)
\sim
\Psi^*_f(x,y,0)\!\int^{a}_{-a}\!dz\,\frac{\hbar}{i}
\frac{\partial}{\partial z}\Psi_i(x,y,z) = 0$,
so that the photoemission matrix element becomes 
susceptible only to the in-plane component of the vector potential. 
The small photoelectron kinetic energy 
$E_{\rm kin}$ achieved by the 7-eV laser is 
favorable for satisfying condition (\ref{eq3}), 
since $\lambda[{\rm \AA}] \sim 12/\sqrt{E_{\rm kin}[\rm{eV}]}$. 
Even when $\lambda$ becomes comparable to $a$, $\Psi_i(z)$ is usually 
an oscillating function for 
$|z| \lesssim a$, and, therefore, the matrix element 
has little dependence on the $\hat{p}_z$ component.

\begin{figure}[htb]
\begin{center}
\includegraphics[width=8.6cm]{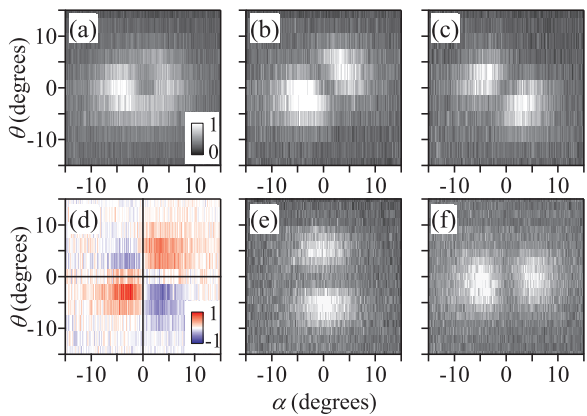}
\caption{\label{fig3} Spectral intensity distributions 
($|E_B| \leq 5$ meV) of SrTiO$_3$:Nb recorded by various polarizations. 
(a) $(I^+ + I^-)/2$. (b) $I^+$. (c) $I^-$. 
(d) $I^D$ showing sign change around $\alpha = 0^{\circ}$ 
(vertical node) and $\theta = 0^{\circ}$ (horizontal node). 
(e) $I^s$. (f) $I^p$. 
}
\end{center}
\end{figure}

{\it The case for SrTiO$_3$:Nb}.--- SrTiO$_3$ is an 
oxide semiconductor having a cubic perovskite structure. 
The bulk can be doped with carriers by incorporating Nb. 
Recently, it has been revealed that 
an inversion layer occurs at the surface of 
semiconducting SrTiO$_3$ independent of 
the carrier concentration of the bulk \cite{Syro, Non}.

We find that the 2D electron gas formed in the inversion layer of 
SrTiO$_3$ is an ideal case that exhibits the horizontal 
and vertical nodes.  
In Fig.\ \ref{fig3}, we show the angular distribution of the spectral weight 
near $E_F$ recorded on a 
(001) surface of 1\%-Nb-doped SrTiO$_3$ annealed in vacuum for 
40 min at 550$^{\circ}$C. Here, 
[100] and [010] are aligned to $\bm{e}_x$ and $\bm{e}_y$, respectively, 
within 5$^{\circ}$. 
A circular Fermi surface can be observed in the $I^+ + I^-$ mapping (a), 
and $I^+$ (b) appears to be a reflection of $I^-$ (c) 
about both the $\theta$ and $\alpha$ axes. Therefore, 
horizontal and vertical nodes occur in $I^D$ (d). The vertical node 
can be understood as a result of the (100) mirror plane 
matching the incidence plane, and the horizontal node can be understood 
from the facts that 
the (010) mirror plane is vertical to the incidence plane and the 
electronic state is 2D. 
The normalized dichroic asymmetry $I^D/(I^+ + I^-)$ 
is a maximum ($>$60 \%) around $\theta = \pm\alpha$. 
The spectral weight mapped by $s$ ($p$) 
polarization 
is bright (dark) at $\alpha = 0^{\circ}$ 
and dark (bright) at $\theta = 0^{\circ}$ 
[see Figs.\ \ref{fig3}(e) and \ref{fig3}(f)], 
indicating that the states probed by the 7-eV laser 
consist of $d_{xy}$ orbitals having odd parity with 
respect to the reflection at the $x=0$ and $y=0$ planes 
\cite{Cuprates_RMP, Syro, Non}. 
The in-plane orbital character of the initial states may 
further facilitate the condition of the 2D confinement 
to be fulfilled.

\begin{figure}[htbp]
\begin{center}
\includegraphics[width=83mm]{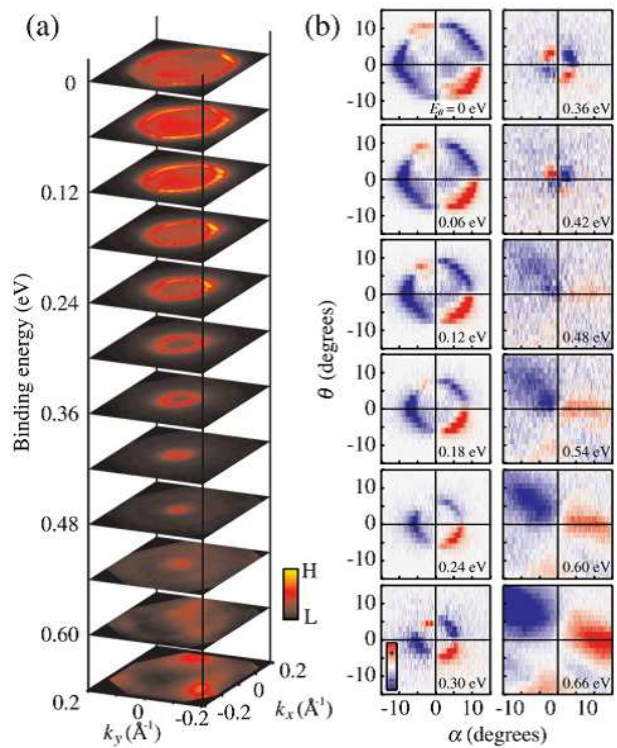}
\caption{CDAD of Cu-doped Bi$_2$Se$_3$. 
(a) Electronic structure recorded by ARPES on the (111) surface 
of Cu$_x$Bi$_2$Se$_3$. Here, $I^+ + I^-$ is 
converted into $k$-space map. The single crystal is 
grown from the melt. (b) $I^D$ at various $E_B$'s. }
\label{fig4} 
\end{center}
\end{figure}

{\it The case for Cu$_x$Bi$_2$Se$_3$}.--- Bi$_2$Se$_3$ is found to be a 
topological insulator \cite{FuKaneMele_PRL, Hasan_Rev} supporting a 
single Dirac-cone dispersion on its 
surface \cite{SCZhang_NPhys, Xia_BiSe_NPhys, Hsieh_Tunable_Spin}. 
Cu intercalation effectively dopes the 
system with electron carriers \cite{Hor_CuBiSe, Wray_NPhys}. 
In the present case, the nominal Cu concentration is $x= 0.17$, and 
$E_F$ is located 480 meV above $E_D$, as shown 
in Fig.\ \ref{fig4}(a). 
The band dispersion in the $k_xk_y$ plane 
($k_x$ is set along $\bar{\Gamma}$-$\bar{M}$ and is 
parallel to $\bm{e}_x$ within 3$^\circ$) 
changes from nearly isotropic 
to hexagonal in going away from $E_D$. 
This can be explained within a 2D $k\cdot p$ Hamiltonian 
constrained under $C_{3v}$ and 
time-reversal symmetry 
\cite{Fu_hexagon_PRL, Kuroda_Hexagon}: 
\begin{eqnarray}
H(k) = v_k(k_x\sigma_y-k_y\sigma_x)
+ k^2/2m^* + \xi/2(k_+^3 + k_-^3)\sigma_z. 
\label{eq4}
\end{eqnarray}
Here, 
$\xi$ is 
responsible for the hexagonal warping, 
$\sigma_i$ is the Pauli matrix, $k_{\pm} = k_x \pm ik_y$, 
$v_k$ contains a $k^2$-order correction, and $1/m^*$ introduces 
particle-hole asymmetry. 
The three-fold pattern observed in the mappings 
at $E_B\ge$ 0.54 eV originates from the bulk valence band, 
and the faint intensity observed inside the hexagon near $E_F$ is 
due to the bulk conduction band \cite{Wray_NPhys}.

In Fig.\ \ref{fig4}(b), we show $I^D$ at various $E_B$'s. 
In the vicinity of $E_D$ (at $E_B$ = 0.42 and 0.36 eV), 
we observe nodal lines at $\theta=0^{\circ}$ and $\alpha=0^{\circ}$. 
This can be explained by noting that 
the effective Hamiltonian [Eq.\ (\ref{eq4})]
up to second order in $k$ is invariant under mirror operations at 
the $x=0$ and $y=0$ planes, and the states therein are 2D, 
so that the conditions for the horizontal and 
the vertical nodes are fulfilled. 
Note that it is the effective Hamiltonian, not the crystal 
surface, that has the mirror symmetry about the $x=0$ plane. 
On the other hand, $I^D$ in the bulk valence-band region at $E_B\ge 0.54$ eV 
does not show the vertical node since the crystal does not have a vertical 
mirror plane. It also does not show the horizontal node 
since the valence-band electronic structure is three-dimensional, 
even though the crystal has a horizontal mirror plane at $y=0$. 

It is apparent that the horizontal node in the topological state 
is gradually distorted in going from $E_D$ to 
$E_F$, even though the crystal as well as the 
effective Hamiltonian has horizontal mirror symmetry. 
This indicates that the topological states lose the condition of 
2D confinement at large $k$, and around $E_F$, 
they penetrate deep into the bulk rather than being 
surface states localized on surface layers.
This is supported by calculations \cite{NJP2010} and 
is reminiscent of a surface-state-to-surface-resonance 
transition with varying 
$k$ observed in the states on metal surface \cite{Al100}. 
The results thus indicate that 
the effective Hamiltonian Eq.\ (\ref{eq4}) is valid only in 
the vicinity of $E_D$. The deviation from Eq.\ (\ref{eq4}) 
may be important 
to understand the possibly exotic superconductivity of Cu$_x$Bi$_2$Se$_3$ 
\cite{Hor_CuBiSe, Wray_NPhys, Ando_PRL2011, FuBerg} 
and the recent observation 
of the topological states away from $E_D$ 
acquiring out-of-plane spin components \cite{Souma}.

In summary, 
we find that the CDADs of both  SrTiO$_3$:Nb around $E_F$ and 
Cu-doped Bi$_2$Se$_3$ around $E_D$ exhibit 
a flower-shaped pattern having horizontal and vertical nodes. 
The vertical node is explained within a well-known geometric effect 
\cite{Schonhense_90}, 
whereas the horizontal node can be understood 
by using a combination of the 
geometry and the 2D character of the initial 
electronic state. The length scale of the 
2D confinement of the initial state 
is set by the de Broglie wavelength of the 
photoelectron final state. 
The horizontal node in CDADs can therefore 
be a measure of the two-dimensionality of the electronic states, 
providing information of 
the penetration depth of surface states and insights into 
the 2D electron gas formed on semiconductor 
surfaces.

This research is supported 
by JSPS through its FIRST Program. 
H.Y.H. and C.B. acknowledge support by the Department of Energy, Office 
of Basic Energy Sciences, Division of Materials Sciences and 
Engineering, under contract DE-AC02-76SF00515.

\end{document}